Evgeny Chutchev

# Some Mathematicians Are Not Turing Machines

## *I. Introduction*

A certain mathematician M, considering some hypothesis H, conclusion C and text P, can arrive at one of the following judgments:

- P does not convince M of the fact that since H, it follows that C (judgment of the type "Not Proved").
- P is the proof that since H, it follows that C (judgment of the type "Proved").

Let us examine the question: Is it possible to replace such a mathematician with an arbitrary Turing machine? In order to answer this question soundly, we shall formalize the concepts of "hypothesis", "conclusion" and "proof" and show that the answer to the question is negative under the two following conditions:

1. M is faultless, namely his judgment "Proved" always implies that since H, it actually follows that C.
2. M recognizes a certain $\mathcal{P}$ as the correct proof of the fact that for certain $\mathcal{H}$ and $\mathcal{C}$, if $\mathcal{H}$, then $\mathcal{C}$ (where $\mathcal{P}$, $\mathcal{H}$, and $\mathcal{C}$ are stated below).

The reason for negative answer lies in the contents of $\mathcal{H}$ and $\mathcal{C}$, whose sense can be expressed informally in the following words:

- $\mathcal{H} =$ "Mathematician M, who operates as Turing machine, is faultless"
- $\mathcal{C} =$ "Mathematician M, who operates as Turing machine, cannot recognize any proof that since $\mathcal{H}$, it follows that $\mathcal{C}$".

The rest of the paper is organized as follows:

Sections II and III give definitions and introduce notations.
Sections IV and V contain the proof of the main statement.
Section VI provides examples related to the concept of "faultless mathematician".
Section VII presents the conclusion.
Section VIII contains a list of definitions and notations.

## *II. General definitions and notations*

For the purposes of formalization, declared in Section I, we introduce the following definitions and notations.

### II.1. Alphabets and symbols

1. Suppose $\mathbb{A}$ is a finite alphabet and symbol $\theta$ does not belong to $\mathbb{A}$. By $\mathbb{A}^*$ denote all words over $\mathbb{A}$. Put $\mathbb{B} \overset{\text{def}}{=} \mathbb{A} \cup \{\theta\}$, and denote by $\mathbb{B}^*$ the set of all words over $\mathbb{B}$. We assume that all expressions, mentioned in this paper, are expressions over the alphabet $\mathbb{B}$.
2. For $\mathbb{V}\subseteq\mathbb{B}^*$ and $\mathbb{U}\subseteq\mathbb{B}^*$, put $\mathbb{V}\theta\mathbb{U} \overset{\text{def}}{=} \{v\theta u: v \in \mathbb{V}, u \in \mathbb{U}\}$.

### II.2. Turing machines

1. We consider Turing machines (TMs) with the tape alphabet $\mathbb{B}$ and denote by $\mathbb{M}$ the set of all TMs, assuming $\mathbb{M} \subset \mathbb{A}^*$.
2. For TM $\mathfrak{M}$ and $b \in \mathbb{B}^*$, denote by $[\mathfrak{M}, b]$ the situation where TM $\mathfrak{M}$ starts on the initial contents of the tape equal to $b$. We write $[\mathfrak{M}, b] \neq \infty$, if $[\mathfrak{M}, b]$ will eventually halt. Otherwise, we write $[\mathfrak{M}, b] = \infty$.

### II.3. Enumerable sets

1. According to the conventional terminology, we call the set $\mathbb{E} \subseteq \mathbb{B}^*$ enumerable, if there exists an arbitrary TM $\mathfrak{E}_{\mathbb{E}}$, such that $\forall_x\big((x \in \mathbb{E}) \Leftrightarrow ([\mathfrak{E}_{\mathbb{E}}, x] \neq \infty)\big)$.
   Note the following:
   - The sets $\mathbb{A}^*$, $\mathbb{B}^*$, and $\mathbb{M}$ are enumerable.
   - If sets $\mathbb{V}\subseteq\mathbb{A}^*$ and $\mathbb{U}\subseteq\mathbb{B}^*$ are enumerable, then the set $\mathbb{V}\theta\mathbb{U}$ is enumerable.
2. For some symbol $\mathfrak{M}$ and enumerable set $\mathbb{E}$, denote by $\mathfrak{M}|_{\mathbb{E}}$ the description of how TM $\mathfrak{M}$ works in situations $[\mathfrak{M}, b]$ for all $b \in \mathbb{E}$. In this case, the general description of $\mathfrak{M}$ is as follows:
   - Read the word $b \in \mathbb{B}^*$, recorded at the beginning of the tape.

- Start up $[\mathfrak{E}_{\mathbb{E}}, b]$.
- If $\mathfrak{E}_{\mathbb{E}}$ halts, start up $[\mathfrak{M}, b]$.

Note that $\forall_{x \notin \mathbb{E}}([\mathfrak{M}, x] = \infty)$ for TM $\mathfrak{M}$ given by $\mathfrak{M}|_{\mathbb{E}}$.

II.4. Special Turing machines

1. Along with TMs, introduced in Subsection II.2.1, we consider special TMs (STMs). The only difference between TM and STM is that STM has two final states, one marked with “Proved” and another marked with “Not Proved”.
2. With each STM $\mathfrak{S}$ we associate TM $\dot{\mathfrak{S}}$, which is obtained by replacing the final state of $\mathfrak{S}$, marked with "Not Proved", with an infinite loop.

II.5. Abridged notations

1. An expression of the type “TM $\mathfrak{M}$ can be constructed” will mean that in the proof of assertion the algorithm of constructing is given.
2. In constructions of the form $\overset{\text{Text}}{\Longrightarrow}$ and $\overset{\text{Text}}{\Longleftrightarrow}$ the “Text” will serve as the substantiation of the corresponding implication or equivalence. Moreover, if “Text” includes a certain designation $d$ within the framework $\left(\boxed{d}\right)$, then this fragment of the “Text” should be read as “Owing to the definition of $d$”.

## *III. Theorems, mathematicians, and provability*

III.1. Definitions

We introduce the following definitions, which formalize the actions of mathematicians.

1. Definitions of sets:
   - By abstract theorem (AT) denote any element of the set $\mathbb{T} = \mathbb{H}\theta\mathbb{C}$, where $\mathbb{H}$ and $\mathbb{C}$ are stated below.
   - By hypothesis of AT denote any element of the set $\mathbb{H} = \mathbb{A}^*$.
   - By conclusion of AT denote any element of the set $\mathbb{C} = \mathbb{A}^*$.
   - By proof of AT denote any element of the set $\mathbb{P} = \mathbb{B}^*$.

   We use the symbols $\mathbb{H}$, $\mathbb{C}$, and $\mathbb{P}$ to emphasize that the set $\mathbb{A}^*$ (or the set $\mathbb{B}^*$) is considered as the carrier, correspondingly, of hypotheses, conclusions, or proofs of ATs.
2. By abstract mathematician (AM) denote any map $\mu\colon \mathbb{T}\theta\mathbb{P} \to \{0, 1\}$, assuming that for $t \in \mathbb{T}$ and $p \in \mathbb{P}$ the equality $\mu(t\theta p) = 1$ means that $\mu$ recognizes the word $p$ as the proof of $t$.
3. With every TM $\mathfrak{M}$ associate the AM $\widehat{\mathfrak{M}}$ according to the following rule:
$$\forall_{t\in\mathbb{T}}\forall_{p\in\mathbb{P}}\left(\left(\widehat{\mathfrak{M}}(t\theta p) = 1\right) \Leftrightarrow ([\mathfrak{M}, t\theta p] \neq \infty)\right).$$
4. Extending the concepts from Subsection II.2.2 to STM, associate with each STM $\mathfrak{S}$ the AM $\breve{\mathfrak{S}}$ according to the following rule:
$$\forall_{t\in\mathbb{T}}\forall_{p\in\mathbb{P}}\left(\left(\breve{\mathfrak{S}}(t\theta p) = 1\right) \Leftrightarrow \begin{pmatrix}\text{In situation } [\mathfrak{S}, t\theta p] \text{ STM } \mathfrak{S} \text{ will reach} \\ \text{the final state marked with “Proved”}\end{pmatrix}\right).$$
5. We consider:
   - Word $p \in \mathbb{P}$ as the proof of AT $t$ for AM $\mu$ and denote this by $p \overset{\mu}{\to} t$, if $\mu(t\theta p) = 1$.
   - AT $t$ as provable by AM $\mu$ and denote this by $\mu \vdash t$, if $\exists_{p\in\mathbb{P}}(p \overset{\mu}{\to} t)$.

III.2. Faultless mathematicians

To introduce the concept of faultless AM we assume that every $a \in \mathbb{A}^*$ has one of the following options:

- Word $a$ is a true statement (in short notation, $a = \mathfrak{t}$).
- Word $a$ is a false statement (in short notation, $a = \mathfrak{f}$).
- Word $a$ can be neither true nor false (in short notation, $a = \mathfrak{u}$), that is, either $a$ is not a statement or, being a statement, contains an internal contradiction.

Using the notation of the form $a = \mathfrak{x}$ for a specific word $a$, we assume that the meaning of such a word is determined by the context of this paper.

Now we call an AM $\mu$ faultless if $\Phi_\mu = \mathfrak{t}$, where
$$\Phi_\mu \overset{\text{def}}{=} \forall_{p\in\mathbb{P}}\forall_{t\in\mathbb{T}}\left(\left(p \overset{\mu}{\to} t\right) \Rightarrow \left((H_t = \mathfrak{t}) \Rightarrow (C_t = \mathfrak{t})\right)\right) \qquad \text{(III.2.1)}$$

and $t = H_t \theta C_t$ for $t \in \mathbb{T}$.

Examples of TM $\mathfrak{M}$, for which $\Phi_{\widehat{\mathfrak{M}}} = \mathfrak{t}$, $\Phi_{\widehat{\mathfrak{M}}} = \mathfrak{f}$, and $\Phi_{\widehat{\mathfrak{M}}} = \mathfrak{u}$, are given in Section VI.

## *IV. Theorem*

The main statement of this paper is the corollary of the theorem proved below. To formulate the theorem and its corollary we denote by $\sigma_{\mathfrak{K}}$ the statement $[\mathfrak{K}, \mathfrak{K}] = \infty$ for an arbitrary TM $\mathfrak{K}$. To prove the theorem and its corollary we rely on the assertion $\forall_{\mathfrak{K} \in \mathbb{M}} (\sigma_{\mathfrak{K}} \neq \mathfrak{u})$.

THEOREM. *For every TM $\mathfrak{M}$, a certain TM $\mathfrak{L}^{(\mathfrak{M})}$, for which $\left(\Phi_{\widehat{\mathfrak{M}}} = \mathfrak{t}\right) \Rightarrow \left(\sigma_{\mathfrak{L}^{(\mathfrak{M})}} = \mathfrak{t}\right)$, can be constructed.*

To prove the theorem we shall prove first the following lemmas.

LEMMA 1. *For every TM $\mathfrak{T}$, a certain TM $\mathfrak{L}_{\mathfrak{T}}$, for which $\left(\sigma_{\mathfrak{L}_{\mathfrak{T}}} = \mathfrak{f}\right) \Leftrightarrow \left(\left[\mathfrak{T}, \sigma_{\mathfrak{L}_{\mathfrak{T}}}\right] \neq \infty\right)$, can be constructed.*

PROOF. Describe TM $\mathfrak{L} = \mathfrak{L}_{\mathfrak{T}}$ by means of $\mathfrak{L}|_{\mathbb{M}}$: in situation $[\mathfrak{L}, \mathfrak{D}]$, where $\mathfrak{D} \in \mathbb{M}$, TM $\mathfrak{L}$ simulates $[\mathfrak{T}, \sigma_{\mathfrak{D}}]$. Then $(\sigma_{\mathfrak{L}} = \mathfrak{f}) \overset{\boxed{\sigma_{\mathfrak{L}}}}{\Leftrightarrow} ([\mathfrak{L}, \mathfrak{L}] \neq \infty) \overset{\boxed{\mathfrak{L}}}{\Leftrightarrow} ([\mathfrak{T}, \sigma_{\mathfrak{L}}] \neq \infty)$.

LEMMA 2. *A certain TM $\mathfrak{N}$, for which $\forall_{\mathfrak{M} \in \mathbb{M}} \forall_{t \in \mathbb{T}} \left(([\mathfrak{N}, \mathfrak{M}\theta t] \neq \infty) \Leftrightarrow \left(\widehat{\mathfrak{M}} \vdash t\right)\right)$, can be constructed.*

PROOF. Denote by $\mathbb{N}$ the set of all natural numbers (including zero). We will not make any distinction between the elements of $\mathbb{N}$ and their notations over the alphabet $\mathbb{B}$, assuming $\mathbb{N} = \mathbb{B}^*$. Now describe TM $\mathfrak{N}$ by means of $\mathfrak{N}|_{\mathbb{M}\theta\mathbb{T}}$:

1. The algorithm of $[\mathfrak{N}, \mathfrak{M}\theta t]$, where $\mathfrak{M} \in \mathbb{M}$ and $t \in \mathbb{T}$, is divided into steps, enumerated by elements of $\mathbb{N}$.
2. At the $k$th step $\mathfrak{N}$ simulates $k - p + 1$ cycles of $[\mathfrak{M}, t\theta p]$ for $p = 0, \dots, k$.
3. If in the process of simulation one of the models has reached a final state, $\mathfrak{N}$ halts. Otherwise, $\mathfrak{N}$ proceeds to the next step of its algorithm.

Then $([\mathfrak{N}, \mathfrak{M}\theta t] \neq \infty) \overset{\boxed{\mathfrak{N}}}{\Leftrightarrow} \left(\exists_{p \in \mathbb{P}} ([\mathfrak{M}, t\theta p] \neq \infty)\right) \overset{\boxed{\overset{\widehat{\mathfrak{M}}}{\to}}}{\Leftrightarrow} \left(\exists_{p \in \mathbb{P}} \left(p \overset{\widehat{\mathfrak{M}}}{\to} t\right)\right) \overset{\boxed{\vdash}}{\Leftrightarrow} (\widehat{\mathfrak{M}} \vdash t)$.

LEMMA 3. *For every TM $\mathfrak{M}$ and every hypothesis $h \in \mathbb{H}$, a certain TM $\mathfrak{T}_{\mathfrak{M},h}$, for which*

$$\forall_{c \in \mathbb{C}} \left(\left(\widehat{\mathfrak{M}} \vdash h\theta c\right) \Leftrightarrow \left(\left[\mathfrak{T}_{\mathfrak{M},h}, c\right] \neq \infty\right)\right),$$

*can be constructed.*

PROOF. Describe TM $\mathfrak{T} = \mathfrak{T}_{\mathfrak{M},h}$ by means of $\mathfrak{T}|_{\mathbb{C}}$: in situation $[\mathfrak{T}, c]$, where $c \in \mathbb{C}$, TM $\mathfrak{T}$ starts up $[\mathfrak{N}, \mathfrak{M}\theta h\theta c]$. Then for every $c \in \mathbb{C}$ we have

$$(\widehat{\mathfrak{M}} \vdash h\theta c) \overset{\boxed{\mathfrak{N}}}{\Leftrightarrow} ([\mathfrak{N}, \mathfrak{M}\theta h\theta c] \neq \infty) \overset{\boxed{\mathfrak{T}}}{\Leftrightarrow} ([\mathfrak{T}, c] \neq \infty).$$

LEMMA 4. *For every TM $\mathfrak{M}$ and every hypothesis $h \in \mathbb{H}$, a certain TM $\mathfrak{L}_{\mathfrak{M},h}$, for which*

$$\left(\sigma_{\mathfrak{L}_{\mathfrak{M},h}} = \mathfrak{f}\right) \Leftrightarrow \left(\widehat{\mathfrak{M}} \vdash h\theta\sigma_{\mathfrak{L}_{\mathfrak{M},h}}\right),$$

*can be constructed.*

PROOF. By the algorithm from the proof of Lemma 3 construct for $\mathfrak{M}$ and $h$ TM $\mathfrak{T} = \mathfrak{T}_{\mathfrak{M},h}$. Thereafter by the algorithm from the proof of Lemma 1 construct for $\mathfrak{T}$ TM $\mathfrak{L} = \mathfrak{L}_{\mathfrak{T}}$. Then

$$(\sigma_{\mathfrak{L}} = \mathfrak{f}) \overset{\boxed{\mathfrak{L}}}{\Leftrightarrow} ([\mathfrak{T}, \sigma_{\mathfrak{L}}] \neq \infty) \overset{\boxed{\mathfrak{T}}}{\Leftrightarrow} (\widehat{\mathfrak{M}} \vdash h\theta\sigma_{\mathfrak{L}}).$$

PROOF OF THE THEOREM. For TM $\mathfrak{M}$ and hypothesis $H = \Phi_{\widehat{\mathfrak{M}}}$, construct by the algorithm from the proof of Lemma 4 TM $\mathfrak{L}^{(\mathfrak{M})} = \mathfrak{L}_{\mathfrak{M},H}$ and introduce the notation $\sigma = \sigma_{\mathfrak{L}^{(\mathfrak{M})}}$. Then

$$(\sigma = \mathfrak{f}) \xRightarrow{\text{Lemma 4}} (\widehat{\mathfrak{M}} \vdash H\theta\sigma) \overset{\boxed{\vdash}}{\Leftrightarrow} \exists_{p \in \mathbb{P}} \left(p \overset{\widehat{\mathfrak{M}}}{\to} H\theta\sigma\right).$$

Thus for $\sigma = \mathfrak{f}$ and $\pi$, for which $\pi \overset{\widehat{\mathfrak{M}}}{\rightarrow} H\theta\sigma$,

$$(H = \mathfrak{t}) \overset{\boxed{H}}{\Rightarrow} \left(\left(\pi \overset{\widehat{\mathfrak{M}}}{\rightarrow} H\theta\sigma\right) \Rightarrow \big((H = \mathfrak{t}) \Rightarrow (\sigma = \mathfrak{t})\big)\right) \overset{\boxed{\pi}}{\Rightarrow} \big((H = \mathfrak{t}) \Rightarrow (\sigma = \mathfrak{t})\big) \xRightarrow{\text{Owing to } H=\mathfrak{t}} (\sigma = \mathfrak{t}).$$

## V. *Corollary of the theorem*

To formulate and prove the corollary of the theorem we introduce the following additional definitions:

1. Denote hypothesis and conclusion of the theorem from Section IV by *The Theorem*.
2. Denote the proof of *The Theorem*, given in Section IV, by *The Proof* of *The Theorem*.
3. We say that AM $\mu$ recognizes *The Proof* of *The Theorem*, if $A_\mu = \mathfrak{t}$, where $A_\mu$ is denoted by the following chain of denotation:
   - $\mathfrak{L}^{(\mathfrak{M})}$ is TM from the statement of *The Theorem*.
   - $C_{\mathfrak{M}} \overset{\text{def}}{=} \sigma_{\mathfrak{L}^{(\mathfrak{M})}}$.
   - $P_{\mathfrak{M}}$ is the text of *The Proof* of *The Theorem* for TM $\mathfrak{M}$.
   - $T_{\mathfrak{M}} \overset{\text{def}}{=} \Phi_{\widehat{\mathfrak{M}}}\theta C_{\mathfrak{M}}$.
   - $A_\mu \overset{\text{def}}{=} \forall_{\mathfrak{M}\in\mathbb{M}} \left(P_{\mathfrak{M}} \overset{\mu}{\rightarrow} T_{\mathfrak{M}}\right)$.

COROLLARY OF *The Theorem*. $\forall_{\text{AM }\mu}\Big(\big(\Phi_\mu = \mathfrak{t}\big)\&\big(A_\mu = \mathfrak{t}\big) \Rightarrow \forall_{\text{STM }\mathfrak{S}}\big(\mu \neq \breve{\mathfrak{S}}\big)\Big)$.

PROOF. Consider AM $\mu$, for which $\big(\Phi_\mu = \mathfrak{t}\big)\&\big(A_\mu = \mathfrak{t}\big)$, and assume $\mu = \breve{\mathfrak{S}}$ for some STM $\mathfrak{S}$. Then $\mu = \widehat{\mathfrak{R}}$, where $\mathfrak{R} = \dot{\mathfrak{S}}$, and

$$\big(\sigma_{\mathfrak{L}^{(\mathfrak{R})}} = \mathfrak{t}\big) \xRightarrow{\boxed{\mathfrak{L}^{(\mathfrak{R})}}\text{ and Lemma 4}} \big(\widehat{\mathfrak{R}} \nvdash \Phi_{\widehat{\mathfrak{R}}}\theta\sigma_{\mathfrak{L}^{(\mathfrak{R})}}\big) \xRightarrow{\boxed{C_{\mathfrak{R}}}\text{ and }\boxed{T_{\mathfrak{R}}}} \big(\widehat{\mathfrak{R}} \nvdash T_{\mathfrak{R}}\big) \overset{\boxed{\vdash}}{\Rightarrow} \left(P_{\mathfrak{R}} \overset{\widehat{\mathfrak{R}}}{\nrightarrow} T_{\mathfrak{R}}\right) \overset{\boxed{A_{\widehat{\mathfrak{R}}}}}{\Longrightarrow} (A_{\widehat{\mathfrak{R}}} \neq \mathfrak{t}).$$

The latter contradicts the assumptions $A_\mu = \mathfrak{t}$ and $\mu = \widehat{\mathfrak{R}}$. Thus, $\sigma_{\mathfrak{L}^{(\mathfrak{R})}} = \mathfrak{f}$, whence by *The Theorem* $\Phi_{\widehat{\mathfrak{R}}} \neq \mathfrak{t}$. Then $\Phi_\mu \neq \mathfrak{t}$, from where $\mu$ fails to satisfy the assumption $\mu = \breve{\mathfrak{S}}$.

## VI. *Examples*

We shall give, as illustration, examples of TM $\mathfrak{M}$, for which $\Phi_{\widehat{\mathfrak{M}}} = \mathfrak{t}$, $\Phi_{\widehat{\mathfrak{M}}} = \mathfrak{f}$, and $\Phi_{\widehat{\mathfrak{M}}} = \mathfrak{u}$.

VI.1. Example of $\Phi_{\widehat{\mathfrak{M}}} = \mathfrak{t}$

Consider an arbitrary TM $\mathfrak{A}$, for which $\forall_{b\in\mathbb{B}^*}([\mathfrak{A}, b] = \infty)$. The definition of $\mathfrak{A}$ implies $\Phi_{\widehat{\mathfrak{A}}} = \mathfrak{t}$. Note that $A_{\widehat{\mathfrak{A}}} = \mathfrak{f}$.

VI.2. Example of $\Phi_{\widehat{\mathfrak{M}}} = \mathfrak{f}$

Consider an arbitrary TM $\mathfrak{B}$, for which $w = \mathfrak{t}$, where $w = \forall_{b\in\mathbb{B}^*}([\mathfrak{B}, b] \neq \infty)$. We have $\left(\left(p \overset{\widehat{\mathfrak{B}}}{\rightarrow} w\theta\neg w\right) \Rightarrow \big((w = \mathfrak{t}) \Rightarrow (\neg w = \mathfrak{t})\big)\right) = \mathfrak{f}$ for every $p \in \mathbb{P}$, hence $\Phi_{\widehat{\mathfrak{B}}} = \mathfrak{f}$. Note that $A_{\widehat{\mathfrak{B}}} = \mathfrak{t}$.

VI.3. Example of $\Phi_{\widehat{\mathfrak{M}}} = \mathfrak{u}$

We shall construct TM $\mathfrak{C}$, for which $\Phi_{\widehat{\mathfrak{C}}} = \mathfrak{u}$, i.e. for which the statement "$\widehat{\mathfrak{C}}$ is faultless" has no sense (can be neither true nor false). For this purpose describe TM $\mathfrak{C}$ by means of $\mathfrak{C}|_{\mathbb{H}\theta\mathbb{C}\theta\mathbb{P}}$ as follows: in situation $[\mathfrak{C}, h\theta c\theta p]$, where $h \in \mathbb{H}$, $c \in \mathbb{C}$, and $p \in \mathbb{P}$:

1. TM $\mathfrak{C}$, sequentially for all TMs $\mathfrak{M}$, constructs the word $\Phi_{\widehat{\mathfrak{M}}}$ and if the equality $h = \Phi_{\widehat{\mathfrak{M}}}$ holds, $\mathfrak{C}$ proceeds to the next item.
2. For TM $\mathfrak{M}$, for which $h = \Phi_{\widehat{\mathfrak{M}}}$ holds, $\mathfrak{C}$ constructs, according to the proof of Lemma 4, the word $C_{\mathfrak{M}}$ and if $c = C_{\mathfrak{M}}$ holds, $\mathfrak{C}$ proceeds to the next item, otherwise $\mathfrak{C}$ enters an infinite loop.
3. TM $\mathfrak{C}$ checks the equality $p = P_{\mathfrak{M}}$ and if it holds, $\mathfrak{C}$ halts, otherwise $\mathfrak{C}$ enters an infinite loop.

According to the proof of the corollary of *The Theorem*,

$$\forall_{\text{AM }\mu}\Big(\big(\Phi_\mu = \mathfrak{t}\big)\&\big(A_\mu = \mathfrak{t}\big) \Rightarrow \forall_{\text{TM }\mathfrak{R}}\big(\mu \neq \widehat{\mathfrak{R}}\big)\Big).$$

But $A_{\widehat{\mathfrak{C}}} = \mathfrak{t}$ (due to the definition of $\mathfrak{C}$), whence $\Phi_{\widehat{\mathfrak{C}}} \neq \mathfrak{t}$.

Further,

$$\left(\Phi_{\widehat{\mathfrak{T}}}=\mathfrak{f}\right)\overset{\boxed{\Phi_{\widehat{\mathfrak{T}}}}}{\Longrightarrow}\exists_{p\in\mathbb{P}}\exists_{h\in\mathbb{H}}\exists_{c\in\mathbb{C}}\left(\neg\left(\left(p\overset{\widehat{\mathfrak{T}}}{\to}h\theta c\right)\Rightarrow\left((h=\mathfrak{t})\Rightarrow(c=\mathfrak{t})\right)\right)\right)\overset{\boxed{\mathfrak{T}}}{\Leftrightarrow}$$

$$\exists_{\mathfrak{M}\in\mathbb{M}}\left(\neg\left(\left(\Phi_{\widehat{\mathfrak{M}}}=\mathfrak{t}\right)\Rightarrow\left(C_{\mathfrak{M}}=\mathfrak{t}\right)\right)\right),$$

but the latter contradicts *The Theorem*, therefore $\Phi_{\widehat{\mathfrak{T}}}\neq\mathfrak{f}$ and, as a result, $\Phi_{\widehat{\mathfrak{T}}}=\mathfrak{u}$.

Note that equality $\Phi_{\widehat{\mathfrak{T}}}=\mathfrak{u}$ is the corollary of the fact that the right side of (III.2.1), for $\mu=\widehat{\mathfrak{T}}$, depends on the truth of $\Phi_{\widehat{\mathfrak{T}}}$.

## *VII. Conclusion*

Let us examine in more detail, what model can we offer for mathematician M, who is provided with the text P as a possible proof of the theorem T = (H, C), where H is the hypothesis of the theorem, and C is the conclusion. Mathematician M can make judgment "Not Proved" (M does not recognize P as the proof of T) or "Proved" (M recognizes P as the proof of T), and may change already made judgment. However, if M is human, then in the lifetime of M there will always be some last, and thus final, judgment relative to P and T. Therefore, it is possible to apply to M the model of an abstract mathematician, defined in Subsection III.1.2 as the map of pairs {T, P} to symbols 0 and 1, such that:

- {T, P} maps to 1 if M arrives at the final judgment "Proved".
- {T, P} maps to 0 otherwise (M arrives at the final judgment "Not Proved", expresses no judgment, does not at all examine the pair {T, P}).

Let us now consider how we can understand the equivalence of mathematician M and certain Turing machine. In the context of examination of pairs {T, P} this machine has to read the texts T and P and, according to the results of its work, it either reports "Proved" or "Not Proved", or makes no report (the latter corresponds to the situation where M makes no judgment). However, exactly that kind of Turing machine we have introduced in Subsection III.1.4 as a possible way of generating an abstract mathematician.

Considering in the introduction (see Section I) the faultless mathematicians, we have involuntarily established the possibility that the concept of truth is suitable for some expressions. Indeed, we have named the mathematician M faultless, if his judgment "Proved" implies that since H, it actually follows that C, i.e. H can be true, C can be true and the truth of C actually arises from the truth of H. This informal definition is well harmonizes with the definition of faultless mathematician, introduced in Subsection III.2.

Further, M can recognize or not recognize the proof of the theorem from Section IV (hereinafter *The Proof* and *The Theorem*). In other words, if Turing machine, mentioned in *The Theorem*, is specified, and $\mathcal{H}$ is the hypothesis of *The Theorem* ($\mathcal{H}=\Phi_{\widehat{\mathfrak{M}}}$), $\mathcal{C}$ is the conclusion of *The Theorem* ($\mathcal{C}=\sigma_{\mathfrak{L}(\mathfrak{M})}$), and $\mathcal{P}$ is *The Proof*, then:

- Either M recognizes *The Proof*, that is M arrives at the judgment "Proved" for $\{(\mathcal{H},\mathcal{C}),\mathcal{P}\}$.
- Alternatively, M does not recognize *The Proof*, that is M arrives at the judgment "Not Proved" for $\{(\mathcal{H},\mathcal{C}),\mathcal{P}\}$ or expresses no judgment or does not at all consider *The Theorem*.

Now we have a full opportunity to apply to mathematician M the corollary of *The Theorem* from Section V:

**A faultless mathematician, who recognizes *The Proof* of *The Theorem*, is not a Turing machine.**

## *VIII. Definitions and notations*

In tables given below, the global definitions and notations (terms) are shown with specifying the number of subsection, where the corresponding definition or notation was denoted.

VIII.1. Definitions and notations

| Abridged notations | | Turing machines | | Theorems, mathematicians, and provability | |
|---|---|---|---|---|---|
| $\overset{\text{Text}}{\Longrightarrow}, \overset{\text{Text}}{\Longleftrightarrow}$ | II.2 | $\mathbb{M}$ | II.2.1 | $\mathbb{T}$ | III.1.1 |

VIII.2.<u>Terms</u>